\documentclass[twocolumn,showpacs,amsmath,amssymb,prd]{revtex4}

\usepackage{graphicx}
\usepackage{dcolumn}
\usepackage{bm}
                
\def\be{\begin{equation}}
\def\ee{\end{equation}}  
\def\ba{\begin{eqnarray}}
\def\ea{\end{eqnarray}}
\def\<{\langle}
\def\>{\rangle}

\begin{document}
\input{epsf}

\title{Atmospheric neutrino
oscillations and tau neutrinos in ice}
\author{Gerardo Giordano$^1$, Olga Mena$^{2}$, Irina Mocioiu$^1$}
\affiliation{$^1$Department of Physics, 
Pennsylvania State University, University Park, PA 16802, USA}
\affiliation{$^2$ 
Instituto de F\'{\i}sica Corpuscular, IFIC, CSIC and Universidad de Valencia, Spain}
\begin{abstract}
The main goal of the IceCube Deep Core Array is to search for 
neutrinos of astrophysical origins. Atmospheric neutrinos are commonly considered as a background for these searches. We show here that cascade measurements in the Ice Cube Deep Core Array can provide strong evidence for tau neutrino appearance in atmospheric neutrino oscillations. A careful study of these tau neutrinos is crucial, since they constitute an irreducible background for astrophysical neutrino detection. 
 \end{abstract}


\date{\today}
\maketitle

\section{Introduction}

Over the last decade, a large number of experiments of different types 
have provided strong evidence for neutrino oscillations and thus 
for physics beyond the Standard Model, see Ref.~\cite{GonzalezGarcia:2007ib} 
and references therein. 
 
Cosmic ray interactions in the atmosphere give a natural beam of neutrinos.  
These atmospheric neutrinos in the GeV range have been
used by the Super-Kamiokande (SK) detector to provide evidence for neutrino
oscillations~\cite{Ashie:2005ik}.
 The large size of neutrino telescopes such as AMANDA,
IceCube and KM3NeT makes possible the detection of a large number of
atmospheric neutrino events with a higher energy threshold, $\sim
100$~GeV, even though the neutrino flux decreases rapidly with energy
($\sim E_\nu^{-3}$). Built to detect neutrinos from astrophysical
sources, or from the decay of Weakly Interacting Massive Particles (WIMPs)  
annihilations~\cite{telescopes}, at 
the high detection threshold energies of these ice/water Cherenkov detectors, 
neutrino oscillation effects would be small. 

Recently, a low energy extension of the IceCube detector, 
the Ice Cube Deep Core array (ICDC) has been proposed and deployed~\cite{deepcore}. 
It consists of six densely instrumented strings ($7$~m spacing among optical modules) located in the deep center region of the IceCube detector plus the seven nearest standard IceCube strings. Its goal is to significantly improve the
atmospheric muon rejection and to extend the IceCube neutrino
detection capabilities in the low energy domain, down to muon or cascade
energies as low as $5$~GeV. The instrumented volume is $15$~Mton. 
Such a low threshold array buried deep inside IceCube will open up a new 
energy window on the universe. It will search for neutrinos from sources 
in the Southern hemisphere, in particular, from the galactic center region, 
as well as for neutrinos from WIMP annihilation, as originally motivated. 
In \cite{MMR} we have proposed neutrino oscillation physics as a further
 motivation for building such an array. In particular, we have analyzed 
the sensitivity of ICDC to the neutrino mass hierarchy. 
ICDC can detect up to $100,000$ atmospheric neutrino events per year, 
orders of magnitude beyond the present data sample, providing rich 
opportunities for detailed oscillation studies. In the same spirit 
of Ref.~\cite{MMR}, we concentrate here on the 
neutrino oscillation analysis in ICDC, focusing now on the cascade signal. 
By exploiting the cascade channel, ICDC could firstly provide 
strong evidence for tau neutrino appearance from oscillations of atmospheric
 neutrinos, greatly improving previous SK results on tau neutrino appearance evidence~\cite{sktau}.

Section \ref{sec:osccascades} reviews briefly our present knowledge of 
the neutrino oscillation parameters as well as their expected errors 
from near future facilities. We then describe the cascade analysis details, 
presenting our main results in section \ref{sec:results}.

\section{Neutrino Oscillations and Cascades in ICDC}
\label{sec:osccascades}

Neutrino data from solar, atmospheric, reactor and accelerator
experiments is well understood in terms of three-flavor neutrino
oscillations. Two $\Delta m^2$ values and two (large) mixing angles
are well determined, while the third mixing angle is limited to be
very small. The CP-violating phase ($\delta$) is completely
unconstrained. In addition, the sign of $\Delta m^2_{31}$ is also
unknown.

The best fit {oscillation} parameter values obtained from present data
are \cite{GonzalezGarcia:2010er}:

\ba
|\Delta m^2_{31}|&=&2.5\times 10^{-3} {\rm eV}^2\\
\Delta m^2_{21}&=& 7.6 \times 10^{-5} {\rm eV}^2\\
\sin^22\theta_{23}&=& 1\\
\tan^2\theta_{12}&=& 0.47
\ea
and $\sin^22\theta_{13}\le 0.15$ for $\Delta m^2_{31}=2.5\times 10^{-3}
{\rm eV}^2$. Notice that an extra unknown in the neutrino oscillation
scenario is the octant in which $\theta_{23}$ lies, if $\sin^2 2
\theta_{23}\ne 1$. This has been dubbed in the literature as the
$\theta_{23}$ octant ambiguity.

In the near future, long baseline experiments like MINOS~\cite{minos} 
will improve the current precision on $\Delta m^2_{31}$ and possibly
discover a non-zero value of $\theta_{13}$, if this is close to the
present upper limit. In a few years, reactor experiments like
DoubleChooz~\cite{dchooz}, RENO~\cite{reno} and DayaBay~\cite{dayabay} will provide improved 
sensitivity to $\theta_{13}$. 
This information can be used as input in our analysis,
reducing some of the parameter uncertainties.

In the past, atmospheric neutrinos in the SK detector
have provided evidence for neutrino oscillations and the first
measurements of $|\Delta m^2_{31}|$ and $\sin^22\theta_{23}$~\cite{Ashie:2005ik}, therefore providing compelling evidence for neutrino oscillations 
versus other more exotic phenomena~\cite{exotic}. 
While facing more systematics than accelerator/reactor experiments 
due to the uncertainties in the (natural source) neutrino fluxes, 
atmospheric neutrinos provide great opportunities for exploring 
oscillation physics due to the large range of energies and pathlengths 
that they span. The ICDC detector 
will collect a data sample which is a few orders of 
magnitude larger than the SK experiment and can also measure energy and
directional information, such that many of the systematic errors associated 
with unknown normalizations of fluxes, cross-sections, etc. can be much better understood by using the data itself.
\begin{figure} [t]
\centerline{\epsfxsize=3.2in \epsfbox{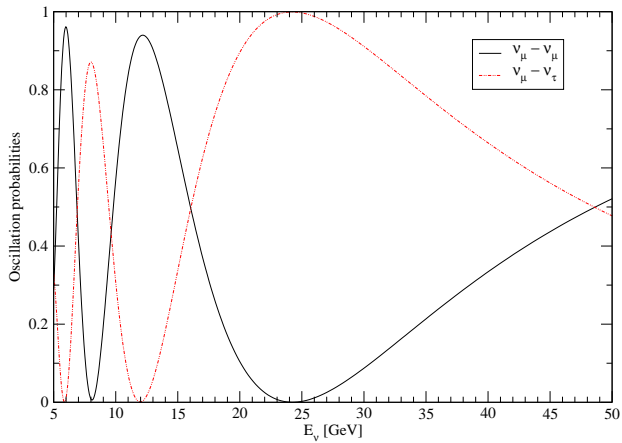}} \caption{$\nu_\mu$
survival probability and $\nu_\mu\to\nu_\tau$ oscillation probability as a function of the neutrino energy (in GeV), assuming upward going neutrinos and $\sin^22\theta_{13}=0.1$.} 
\label{fig:probtau}
\end{figure} \\
The IceCube detector and its Deep Core extension are optimized for detecting muon tracks from the charged current interactions of $\nu_\mu$. It is, however, possible to also detect cascades~\cite{cascades}. While these type of events no longer provide directional information, their energy can be measured quite precisely. We analyze here the cascade events in the ICDC array. There are several contributions to the cascade signal: charged current interactions of $\nu_e$, neutral current interactions of all neutrino flavors and electromagnetic and hadronic decays of tau leptons produced in charged current interactions of $\nu_\tau$. There are several important observations which suggest that the $\nu_\tau$ signal can be significant and can provide evidence for $\nu_\tau$ appearance from oscillations of $\nu_\mu$: 
\\
$\bullet$ First, 
the atmospheric electron neutrino and electron antineutrino fluxes at the relevant energies are significantly lower than the muon neutrino flux, such that the $\nu_e$ charged current interactions do not completely overwhelm the event rate. \\
$\bullet$
In addition, the energy range covered by ICDC corresponds to a maximum of 
$\nu_\mu \to \nu_\tau$ oscillations (minimum of $\nu_\mu$ survival), as can be noticed from Fig.~\ref{fig:probtau}. 

The large flux of atmospheric muon neutrinos can thus lead to a large flux of tau neutrinos. This has already been noted in \cite{MMR}, where it was shown that  $\nu_\mu\to\nu_\tau\to\tau\to\mu$ provides a non-negligible contribution to the muon track rate. It is also important to note that, unlike for the SK detector, which is sensitive at much lower energies, tau threshold production effects are relatively small, only affecting the lowest energy events detected by ICDC.

We investigate the neutrino energy range between $10$~GeV and $100$~GeV, assuming bins of $5$~GeV width in the observable energy. Cascades have very little directional information, especially at these low energies, so we integrate over all upward going directions~\footnote{We have integrated over the zenith angle direction $\theta$ of the incident neutrinos, from $c_\nu=-1$ (vertically upward going) up to  $c_\nu=0$ (horizontally incident), being $c_\nu$ the cosine of $\theta$.}. The downward going neutrinos are largely unaffected by oscillations, so they can be used for determining the atmospheric neutrino flux and thus the contribution of the $\nu_e$ charged current interactions to the overall cascade rate. In our numerical calculations we have taken into account full three flavor oscillations. It is however straightforward to see  that solar parameters do not play an important role in the analysis due to the rather high energy threshold of ICDC. Also, $\theta_{13}$ effects, while in principle observable for values of $\theta_{13}$ close to the present bound, do not affect any conclusions regarding $\nu_\tau$ rates, which are determined by the (maximal) atmospheric mixing angle $\theta_{23}$.
 
The spectrum of $\nu_e$ induced events as a function of observable energy is given by:
\begin{widetext}
\be
\frac{{\rm d} N_e}{{\rm d} E_{obs}}= 2\pi n_T t  
{\displaystyle{\int}} d\cos\theta~ V~\sigma^{\rm  CC}_{(\nu_e)}(E_\nu)\left( \displaystyle{\frac{d\phi_{\nu_e}(\theta,E_\nu)}{dE_\nu d\Omega} }~ P_{\nu_e\to\nu_e} (E_\nu,\theta)+\displaystyle{\frac{d\phi_{\nu_\mu}(\theta,E_\nu)}{dE_\nu d\Omega} }~ P_{\nu_\mu\to\nu_e}(E_\nu,\theta)\right)+(\nu\to\bar\nu)\, ,
\label{eq:nuecc}
\ee
\end{widetext}
where $n_T$ is the number density of targets, $V$ is the volume of the detector, $\theta$ is the zenith angle direction of the neutrino and t is the observation time. The detector is a cylinder of $250$~m diameter and $350$~m height, i.e. the total physical mass is of around $15$~Mton. The low energy events however are mostly single-string events and we consider the effective volume in this case to be six cylinders of $40$~m radius centered around the six densely instrumented ICDC strings. 

The second term of Eq.~(\ref{eq:nuecc}), which contains the contribution from oscillations of $\nu_\mu \to \nu_e$, is negligible in practice. This contribution is very small even for the maximum current allowed value of $\sin^22\theta_{13}$ and after considering enhancements in the oscillation probabilities due to matter effects inside the Earth. It could, however,  become relevant in the presence of non-standard neutrino interactions.

For charged current electron neutrino interactions, the entire energy of the incident neutrinos is transferred to the cascade and thus measured.

For neutral current interactions, only a fraction $y$ of the initial neutrino interaction is transferred to the cascade. Due to the steep energy dependence of the atmospheric neutrino fluxes, their contribution is thus expected to be smaller. The spectrum for neutral current (NC) interactions as a function of the observable energy is given by:
\begin{widetext}
\be
\frac{{\rm d} N_{NC}}{{\rm d} E_{obs}}= 2 \pi n_T t  \,
{\displaystyle{\int}} d\cos\theta~ V \displaystyle{\int}_{E_{obs}}^\infty dE_\nu \frac{d\phi_{\nu_i}}{dE_\nu d\Omega}\frac{1}{E_\nu} P_{\nu_i\to \nu_j}(E_\nu,\theta)
\frac{{\rm d}\sigma^{\rm  NC}_j}{{\rm d} y }(y, E_\nu)\displaystyle{\Big|_{y=E_{obs}/E_\nu}}
 + (\nu\to\bar\nu) \, .\ee
\end{widetext}

For tau neutrinos, charged current (CC) interactions lead to two types of contributions to cascade events: electromagnetic and hadronic, depending on the tau decay mode. The rate for hadronic events is given by:
\begin{widetext}
\be
\frac{{\rm d} N_\tau^{\rm had}} {{\rm d} E_{obs}}=2 \pi n_T t  \,
{\displaystyle{\int}} d\cos\theta~ V \displaystyle{\int}_{E_{obs}}^\infty dE_\nu \displaystyle{\int}_{E_{obs}}^{E_\nu} dE_\tau \frac{d\phi_{\nu_\mu}}{dE_\nu d\Omega}P_{\nu_\mu\to\nu_\tau} (E_\nu,\theta)
\frac{{\rm d}\sigma^{\rm  CC}_\tau}{{\rm d} E_\tau }(E_\tau,E_\nu)\frac{{\rm d}n^{\rm had}}{{\rm d} E_\tau}(E_\tau)+(\nu\to\bar\nu)\, .
\label{eq:tauhad}
\ee
\end{widetext} 
For our numerical calculations, we use differential neutrino cross-sections as given by \cite{GQRS} with CTEQ6 parton distribution functions. In terms of dimensionless variables typically used for these cross-sections we have: ${{\rm d}\sigma^{\rm  CC}_\tau}/{{\rm d} E_\tau }\sim 1/E_\nu \, {{\rm d}\sigma^{\rm  CC}_\tau}/{{\rm d} y }$ with $y=1-E_\tau/E_\nu$ and ${{\rm d}n^{\rm had}}/{{\rm d} E_\tau}\sim E_{ons}^2/E_\tau \, {{\rm d}n^{\rm had}}/{{\rm d} z}$ with $z=1-E_{obs}/E_\tau$.
It is important to note that for tau neutrinos, tau threshold suppression is still important in the lower energy range discussed here. We have included the threshold corrections following Ref.~\cite{hallsietau}. The $3.5$~GeV tau lepton production threshold energy has made the tau neutrino detection very difficult up to now, since the atmospheric neutrino flux for energies above tau lepton production threshold is very low for existing detectors. For instance, for the SK experiment,  and assuming maximal mixing in the $\nu_\mu-\nu_\tau$ sector, only one $CC$ tau neutrino event is expected per kton-year of exposure. The ICDC experiment, which benefits from a much larger instrumented volumen than the SK experiment, is in an unique position to detect tau neutrino interactions, as it has good sensitivity in an energy range where neutrino oscillations lead to a large number of tau neutrinos and, at the same time, is high enough to no longer be strongly affected by tau threshold suppression. 

The rate for electromagnetic cascade events is given by an expression similar to Eq.~(\ref{eq:tauhad}), the only difference being in the decay rate ${\rm d}n/{{\rm d} E_\tau}$. 

\section{Results}
\label{sec:results}

\begin{figure} [t]
\centerline{\epsfxsize=3.2in 
\epsfbox{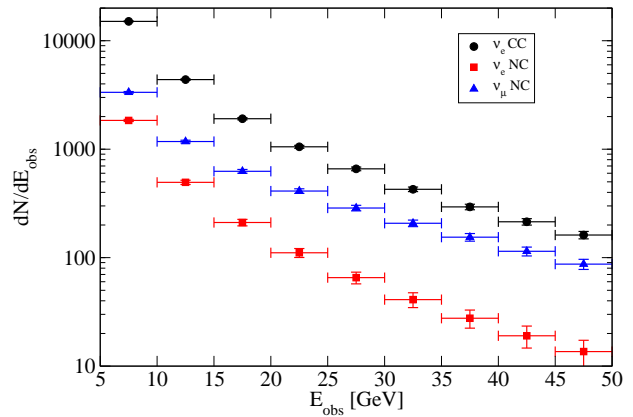}
} 
\caption{Electron CC and all flavors NC events for one year ICDC exposure, see text for details in the calculation.} 
\label{fig:bkgd}
\end{figure}

\begin{figure} [t]
\centerline{\epsfxsize=3.2in \epsfbox{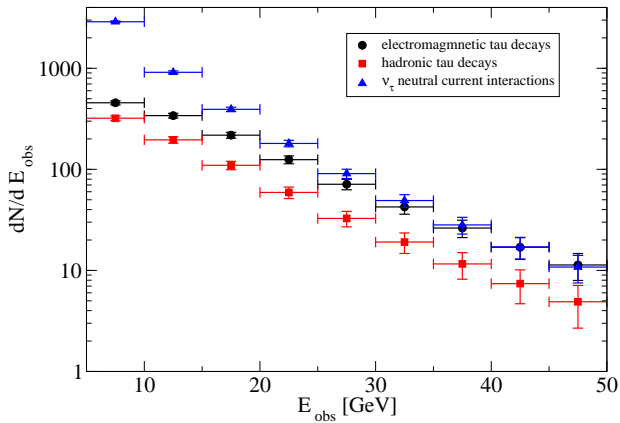}
} 
\caption{Tau cascade events for one year of ICDC exposure.} 
\label{fig:tau}
\end{figure}

Figure \ref{fig:bkgd} shows the cascade rates coming from $\nu_e$ charged current interactions and neutral current interactions of $\nu_e$ and $\nu_\mu$. These events are indistinguishable from the $\nu_\tau$ cascade events that are of interest to us, so they constitute an irreducible background for the $\nu_\tau$ search.
Figure \ref{fig:tau} illustrates the cascade rates from charged current $\nu_\tau$ interactions followed by hadronic or electromagnetic tau decays, as well as $\nu_\tau$ neutral current interactions.

While the background is significant, the number of events is very high. Neglecting systematics errors, the statistical significance could be defined as
\begin{equation}
S=\frac{n_s}{\sqrt{n_s+n_b}}~,
\end{equation}
$n_s$ being the number of $\nu_\tau$ events and $n_b$ the number of background events (from $\nu_e$ changed current interactions and $\nu_{e,\mu}$ neutral current interactions). A statistically significant ($3 \sigma$) $\nu_\tau$ appearance signal could be obtained in only a \emph{few months} of observation.  
However, systematic uncertainties will limit the analysis. 
One of the biggest uncertainties is the energy threshold for cascade 
identification. At the lowest energies considered here, muon track 
events would be very short~\footnote{A $5$~GeV muon will have a track in ice of $\sim 25$~m.} and therefore, cascade events and muon 
track events would be indistinguishable. If cascade events can not be 
distinguished from muon tracks, the background to the $\nu_\tau$ 
signal becomes significantly higher (by about an order of magnitude) 
due to the contribution from the muon event tracks. Even in the case where
 muon track events and cascade events can not be 
distinguished at low energies, the $\nu_\tau$ signal can become statistically 
significant in a year of exposure. Good muon track reconstruction would be extremely useful for reducing 
this systematic uncertainty and might be achievable with the planned additional 
strings in the center of the detector. 
There are other systematic uncertainties affecting the analysis, as the 
knowledge of the interaction cross-sections, atmospheric neutrino fluxes, 
and other neutrino oscillation parameters. However, these systematic errors
are expected to be under control in the next few years, exploiting data 
from future reactor and accelerator experiments as well as atmospheric neutrino data from the ICDC experiment in different angular and energy ranges than the ones used for the $\nu_\tau$ analysis. 
Appearance of tau neutrinos from oscillations of atmospheric $\nu_\mu$ 
is thus likely to be detected in the near future.

\section{Outlook}
\label{sec:outlook}

The IceCube detector and its Deep Core array provide a great
opportunity for studies of atmospheric neutrinos. Being the largest
existing neutrino detector, it will accumulate a huge number of
atmospheric neutrino events over an enormous energy range, thus
allowing for detailed studies of oscillation physics, Earth density,
atmospheric neutrino fluxes and new physics~\cite{concha0}. In order to extract all
this information it is necessary to use energy and angular
distribution information, as well as flavor composition, all possible
to obtain with the IceCube detector. Qualitatively, there are three 
main energy intervals and three main angular regions which are sensitive to 
different types of physics.

At very high energies, above $10$~TeV, neutrino interaction
cross-sections become high enough that neutrinos going through the
Earth start getting attenuated~\cite{concha1}. 
This effect is sensitive to neutrino
interaction cross-sections and to the density profile of the Earth. 

The ``intermediate'' energy region, between $50$~GeV and $1$~TeV can
provide good information about the atmospheric neutrino flux, which
can be used to improve the uncertainties in the simulated atmospheric
neutrino fluxes~\cite{concha2}.

In our paper we concentrated on the ``low'' energy region, below about $40$~GeV, where neutrino oscillation effects can be significant. 
Although the IceCube detector has higher energy thresholds, its low energy 
extension, the Ice Cube Deep Core array (ICDC), extends the IceCube neutrino
detection capabilities in the low energy domain, down to muon or cascade
energies as low as $5$~GeV. We have shown that tau neutrinos from oscillations 
of atmospheric muon neutrinos could be detected in the next few years 
with a high significance level, providing therefore the first sizable sample of tau neutrinos. This large number of tau neutrinos could allow for measurements 
of $\nu_\tau$ interaction cross-sections and studies of non-standard neutrino 
interactions, which are largely unconstrained in the tau sector. 
In summary, ICDC offers a unique window towards a better understanding of 
neutrino properties due to its very high atmospheric neutrino statistics. 
Careful studies of the expected atmospheric neutrino oscillation signals in ICDC, as the one carried out here for the $\nu_\tau$ appearance signal, 
are mandatory, since atmospheric neutrinos constitute an irreducible background to astrophysical neutrino searches.

\section*{Acknowledgments} 

This work was supported in part by the NSF grant PHY-0855529.
I. M. would like to thank the Aspen Center for Physics where part of this work was completed. O.~M. work is supported by the MICINN (Spain) Ram\'on y Cajal contract, AYA2008-03531 and CSD2007-00060.

\end{document}